\documentclass[a4paper,12pt]{article}
\usepackage[utf8x]{inputenc}
\usepackage[lmargin=2cm,rmargin=2.0cm,tmargin=2cm,bmargin=2cm]{geometry}
\usepackage[pdftex]{graphicx}
\usepackage{caption}
\usepackage{subcaption}
\usepackage{color}
\usepackage{amsmath}
\usepackage{amssymb}
\usepackage[numbers]{natbib}
\usepackage{authblk}
\usepackage{textcomp}

\usepackage{caption}

\title{\textbf{Interplay between crystallization and glass transition in binary Lennard-Jones mixtures}}
\author{Atreyee Banerjee$^{1}$, Suman Chakrabarty$^{2}$, Sarika Maitra Bhattacharyya$^{1} \footnote{electronic address : mb.sarika@ncl.res.in} $}
\affil{\textit{$^{1}$Department of Polymer Science and Engineering, CSIR-National Chemical Laboratory, Pune-411008, India}}
\affil{\textit{ $^{2}$Physical Chemistry Division, CSIR-National Chemical Laboratory, Pune-411008, India}}

\date{}
\begin{document}

\maketitle
\begin{abstract}

In this work we explore the interplay between crystallization and glass transition in different binary mixtures by changing their inter-species interaction length and 
also the composition. We find that only those systems which form bcc crystal in the equimolar mixture and whose global structure for 
larger $x_A$ ($x_A=0.6$, where $x_A$ is the mole
 fraction of the bigger particles) is a mixed fcc+bcc phase,
do not crystallize at this higher composition. However, the systems whose equimolar structure is a variant of fcc (NaCl type crystal) and whose global structure at 
larger $x_A$ is a mixed
 NaCl+fcc phase, crystallize  easily to this mixed structure. We find that the stability against crystallization of this ``bcc zone" is due to the frustration between the 
locally preferred 
structure (LPS) and the mixed bcc+fcc crystal. Our study suggests that when the global structure is a mixed crystal where a single species contributes to both the crystal forms
 and where the two crystal forms have large difference in some order parameter related to that species then this induces frustration between the 
LPS and the global structure. This frustration makes the systems good glass former.
 When $x_A$ is further increased ($0.70\leq x_A < 0.90$) the systems show a tendency towards mixed fcc crystal formation. 
However, the ``bcc zone" even for this higher composition is found to be sitting at the bottom of a V shaped phase diagram formed by two different variants of the fcc crystal
 structure, leading to 
its stability against crystallization.

\end{abstract}

\section{Introduction}

The origin of glass transition and the stability of a supercooled liquid against crystallization is 
still not well understood and is an open question \cite{inoue_international,tanaka-epje}. It is usually found that during fast cooling due to a  large change in viscosity,
 crystallization can be avoided and the system is vitrified.  The vitrified materials are tougher, stronger and  have large strain limits.
 When compared to their crystalline counterparts these glassy materials can be easily used to prepare homogeneous, isotropic solids in large dimensions. 
Although vitrification is desirable but not all supercooled systems form glasses, many undergo crystallization. 
Thus it is important to understand the origin of stability of supercooled liquids against crystallization.
 In the metallic glass community some empirical rules are used based on the analysis of glass forming ability (GFA) of metallic alloys \cite{Acta_mater}. 
The rules state that i) the system should have more than three components, ii) the size ratio between the components should be about 12 \%, and iii) the enthalpy 
of mixing should be negative. Although having more than three components is a desirable criteria for GFA but some binary metallic alloys are also known to form 
glasses \cite{inoue_international}. One such glass former Ni$_{80}$P$_{20}$ has been the motivation behind the development of a well known model
 system, known as the Kob-Anderson (KA) model \cite{kob,stillinger}. This model system has been extensively used in computer simulation studies of supercooled liquids \cite{dyre,valdes-jcp}. 
 The  KA model has never been
 found to crystallize except for one case \cite{dyre}. However, the origin of its stability against crystallization is not fully understood 
\cite{ dyre, valdes-jcp,harowell, harrowell-jpcb, harrowell-jcp, doye, vlot}. 

There are some frustration based approaches to explain the stability of supercooled liquids 
\cite{  tanaka-epje,harowell,frank, tarjus, tarjus-2, tarjus-charbo, tanaka-jnoncryst1,tanaka-jnoncryst2, tanaka-nature06,tanaka-vshaped-prl,tanaka22,tanaka23,tanaka29}. 
The role of frustration in supercooling
 has been invoked first  by Frank \cite{ frank}. He has pointed out that the local icosahedral ordering of liquid although cannot be spanned in space, is locally more stable 
than crystal ordering. 
The crystal ordering wins over only because it becomes economical when spanned over long range. Thus when a liquid is cooled it requires substantial costly rearrangement of 
molecules to crystallize and this slows down the crystallization process and promotes supercooling below the melting point. Kivelson  {\it et al }  have
 proposed a frustration
 theory to connect the slow dynamics in the system to the local preferred structure (LPS) \cite{tarjus}. According to their theory, the liquid will prefer to freeze in 
the locally preferred liquid structure (icosahedral for Lennard-Jones (LJ) liquids) which is different from the crystal structure. Since the local structure cannot tile 
the ordinary three dimensional space, in trying to do so the liquid will be geometrically frustrated and will break up into domains.
 The rearrangement in these domains gives rise to the slow dynamics and glass transition.
 A different picture of  frustration has been proposed by Tanaka and co-workers
\cite{ tanaka-epje,tanaka-jnoncryst1,tanaka-jnoncryst2,tanaka-nature06,tanaka-vshaped-prl,tanaka22,tanaka23,tanaka29}. According to their theory, 
liquid-glass 
transition is connected to  crystallization \cite{tanaka-jnoncryst1,tanaka-jnoncryst2}. They have proposed that there is frustration between short 
range bond ordering to form LPS
and long range density ordering which gives rise to crystal structure. This frustration leads to GFA of a system. 
 Thus it is obvious that the origin and the role of frustration are different in all these different studies. 

As mentioned before the KA model has been extensively used to study the dynamics of supercooled liquids because of its stability against crystallization.
 There has been a large number of studies by different groups, devoted to the understanding of the kinetics of 
crystallization \cite{ dyre, valdes-jcp,harowell, harrowell-jpcb}and also the stability of crystal phases \cite{harowell, harrowell-jcp, doye, vlot}.
 These studies have been performed for not only KA model, but in general for the binary LJ mixture. 
Fernandez and Harrowell have performed crystal phase analysis of binary LJ for different inter-species interaction length and also for different 
compositions \cite{harrowell-jcp}. 
Their study has revealed that for the KA model at T=0 the most stable equilibrium structure is a coexistence between AB (CsCl) crystal and pure A (fcc) crystal with a coherent 
(001) interface. They  have also suggested that the crystal growth of KA model might be frustrated because of the competition 
between the growth of AB (CsCl form) and A (fcc form) structures \cite{harowell}. According to them this frustration might be the origin of stability of the KA model.
Valdes {\it et al.}  have studied the binary LJ mixture at different compositions \cite{valdes-jcp}. They claim that for the compositions where the system undergoes amorphization 
they find either CsCl type or fcc-hcp type crystal seeds in the liquid. Thus they predict that since both the structures  do not coexist there is no competition 
between these two type of crystal growth.
Toxvaerd {\it et al.} have pointed out that negative mixing enthalpy or energy leads to the system to be stable supercooled mixtures \cite{dyre}. 
Doye {\it et al.} have done isolated stable cluster analysis of the preferred coordination of A atoms around the smaller B atoms for both $x_A=0.80$ (KA model) 
which is known not to crystallize, and for $x_A=0.50$ which quite easily forms a CsCl type interpenetrating bcc crystal structure \cite{doye}. 
They found that for $x_A=0.80$ the structures are related to the square anti-prism. Similar structures were also found to be present in the local structure 
analysis of supercooled KA liquid \cite{harrowell-jpcb, coslovich-pastore}. 
Doye {\it et al.} have also found that for $x_A=0.50$ stable structure is the CsCl type crystal. 
Equimolar binary LJ systems for different inter-species interaction lengths ($\sigma_{12}$) are also known to crystallize in different forms depending on 
the $\sigma_{12}$ values \cite{vlot}.

In this study we explore the interplay between crystallization and supercooling for a number of binary LJ mixtures at various compositions and inter-species interaction lengths
 described by $s$ where  $s= \sigma_{12} / \sigma_{11}$. We have used the local Bond  Orientational Order (BOO) parameters and the local coordination number to identify
 the locally preferred structures and crystal structures.
Recently the local BOO parameters have been extensively used by Tanaka and coworkers to study properties of not only crystals but also 
supercooled liquids \cite{tanaka-nature12}.
 The systems studied here have all 
negative enthalpy of mixing and the size ratio between the two components are kept fixed at 12$\%$. In this range of systems we have found that some easily
 crystallize and some remain in supercooled liquid state. The focus of this study is to understand: i) the origin of this variation in the crystallization behaviour and 
ii) the origin of stability against crystallization in terms of frustration between two crystal forms and also 
the frustration between the LPS and the global structure.

The simulation details are given in the next section. In section 3 we have the results and discussion, and  section 4 ends with a brief summary.

\section{Simulation Details}
We have performed molecular dynamics study with composition variation and interaction length variation.
 The atomistic models which are simulated are two component mixtures of N=500 classical
particles, where particles of type 
{\it i} interact with those of type {\it j} with pair 
potential, $U_{ij}(r)$,  where r is the distance between the pair. 
$U_{ij}(r)$ is described by a shifted and truncated Lennard-Jones (LJ) potential, as given by:
\begin{equation}
 U_{ij}(r)=
\begin{cases}
 U_{ij}^{(LJ)}(r;\sigma_{ij},\epsilon_{ij})- U_{ij}^{(LJ)}(r^{(c)}_{ij};\sigma_{ij},\epsilon_{ij}),    & r\leq r^{(c)}_{ij}\\
   0,                                                                                       & r> r^{(c)}_{ij}
\end{cases}
\end{equation}

\noindent where $U_{ij}^{(LJ)}(r;\sigma_{ij},\epsilon_{ij})=4\epsilon_{ij}[({\sigma_{ij}}/{r})^{12}-({\sigma_{ij}}/{r})^{6}]$ and
 $r^{(c)}_{ij}=2.5\sigma_{ij}$. Subsequently, we'll denote A and B types of particles by indices 1 and 2, respectively.

The different models are distinguished by different choices of lengths and composition parameters. Length, temperature and
time are given in units of $\sigma_{11}$, ${k_{B}T}/{\epsilon_{11}}$ and $\surd({m\sigma_{11}^2}/{\epsilon_{11}})$, 
respectively.  
Here we have simulated various binary mixtures  
with the interaction parameters  $\sigma_{11}$ = 1.0, $\sigma_{22}$ =0.88 , $\epsilon_{11}$ =1, $\epsilon_{12}$ =1.5,
 $\epsilon_{22}$ =0.5, $m_{1}$ =1, $m_2$=0.5 and the inter-species interaction length $\sigma_{12}$ 
has been varied such that the size ratio $s = \sigma_{12}/\sigma_{11}$ varies from 0.70 to 0.94 with an interval of 0.02. 
 In this article $\sigma_{12}$ and $s$ have been used interchangeably because for  $\sigma_{11}=1$, $s=\sigma_{12}$.
  We have also 
simulated systems with different compositions, varying $x_A$ from 0.50 to 0.90, where $x_A$ is the mole fraction of the bigger A type particles. 
The systems with $s$= 0.94 follow LB rule of mixing for distance \cite{lorentz}. Note that the mixture with $s=0.80$ and $x_A=0.80$ is the 
well-known Kob Anderson (KA) Model which is extensively used as a model supercooled liquid \cite{dyre,valdes-jcp}.

The molecular dynamics (MD) simulations have been carried out using the LAMMPS 
package \cite{lammps}.
We have performed MD simulations in the isothermal−isobaric ensemble (NPT) using  Nos\'{e}-Hoover thermostat and Nos\'{e}-Hoover barostat with integration timestep 0.005$\tau$. The time
constants for  Nos\'{e}-Hoover thermostat and barostat are taken to be 100 and 1000 timesteps, respectively.
The sample is kept in a cubic box with periodic boundary condition. To study crystallization we have done stepwise cooling with $\Delta T^*=0.1$. At each temperature the system has been usually equilibrated for 10ns, however in certain
cases where it was difficult to crystallize we have run the simulation even for 10$\mu s$. 

 Bond  Orientational Order parameter was 
first prescribed by
Steinhardt {\it et al.}  to characterize specific crystalline structures \cite{steinhardt}. Leocmach {\it et al.}  have shown that these BOO parameter can be used not only 
for crystals but also for supercooled liquids where although there is no clear crystalline order but a tendency towards  crystalline ordering can be identified by the transient
 local BOO analysis
 \cite{tanaka-nature12}.

To characterize specific crystal structures and also to identify the tendency towards crystallization in a liquid here  we have calculated the local BOO 
parameters  ($q_{lm}$) of \textit{l}-fold symmetry as a 2\textit{l}+1 vector , 
\begin{eqnarray}
  q_l =\sqrt{\frac{4\pi}{2l+1}\sum_{m=-l}^{l}\arrowvert q_{lm}\arrowvert^2} \nonumber \\
 q_{lm}(i)=\frac{1}{N_i} \sum_{0}^{N_{i}} Y_{lm}(\theta(r_{ij}), \phi(r_{ij})) 
\end{eqnarray}
where $Y_{lm}$ are the spherical harmonics,  $\theta(r_{ij})$ and $\phi(r_{ij})$ are 
spherical coordinates of a bond $r_{ij}$ in a fixed reference frame,
and $N_i$ is the number of neighbours of the {\it i}-th particles. Two particles are considered neighbours if $r_{ij}< r_{max}$, where $r_{max}$ is the first minimum of the radial distribution
function (RDF).



\section{Results}

The range of system studied here have negative mixing enthalpy and the size ratio(s) between the two components are always kept 12\%. 
Crystallization has been identified by a sudden drop in the potential energy while gradually cooling the system. We have further quantified the crystallization 
process by calculating the RDF and the local BOO parameters before and after the energy drop.
We have used the local BOO parameters to identify not only the crystal forms but also the transient ordering present in the liquids. For the range of system studied here,
 it is found that primarily face centered cubic (fcc),
 body centered cubic (bcc), simple cubic (sc) and hexagonal closed packed (hcp) structures are formed . The $q_4$ and $q_6$ parameters for these
 different perfect crystal structures are listed in Table-1, which we have used to identify
our crystal structures. Instead of calculating the average $q_4$ and $q_6$ parameters, we have calculated the probability distribution of these values over individual 
particles and over the length of the trajectory. Such a distribution provides us more microscopic information regarding the tendencies of local structure formation even when
a perfect crystal structure is not achieved.

\begin{table}
\caption{Reduced invariants $q_4$ and $q_6$ for face-centered cubic (fcc), body-centered cubic (bcc), simple cubic (sc) and hexagonal closed packed (hcp) structures. }
\centering
\begin{tabular}{ l r r }
 \hline
\hline
   & $q_4$& $q_6$ \\
 \hline
  
fcc & 0.191&0.575  \\
bcc & 0.036&0.511  \\
sc & 0.764&0.355  \\
hcp & 0.097& 0.485\\
\hline
\hline
\end{tabular}

\end{table}

\begin{figure}[]
\centering
\includegraphics[width=0.5\textwidth]{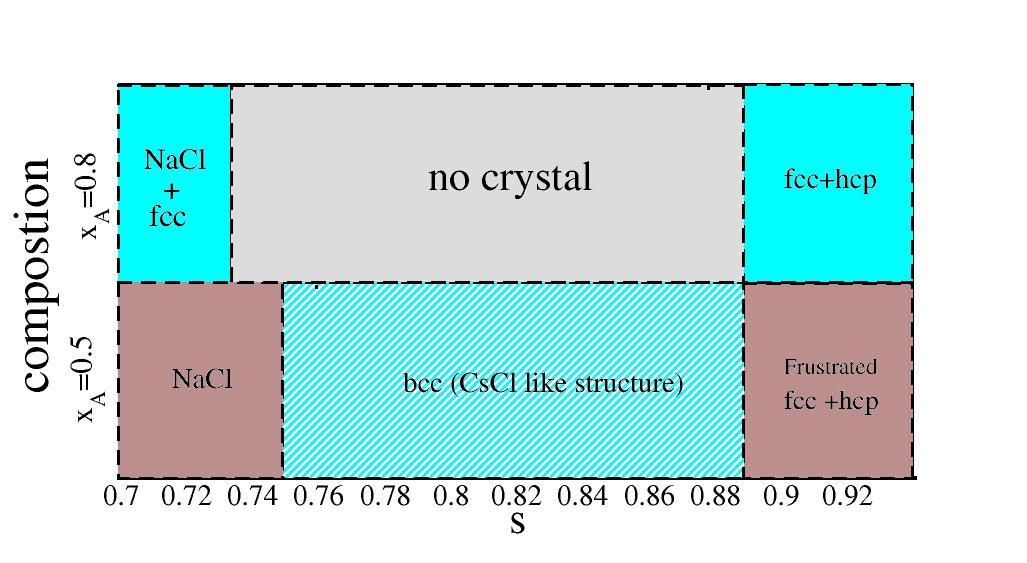}
\caption{ Phase diagram of different types of crystal structures and amorphous structures plotted against variation of interaction lengths ($s$) for 
$x_A=0.50$ and $x_A=0.80$. In the left side the brown zone is where NaCl type of crystal is found and in the right side the brown zone is where the distorted
 hcp + fcc crystal structure is obtained for the $x_A=0.50$.
 At this composition CsCl type bcc structure is found in the intermediate values of s and shown by the dashed cyan zone (\textbf{bcc zone}). 
The above panel of the plot describes different
 structures obtained for $x_A=0.80$. The region forming NaCl+ fcc crystal is shown in the left cyan zone and the one forming fcc + hcp crystal is shown in the right 
cyan zone. We do not find any crystal in the full range of brown zone. At $s=0.74$ and $s=0.90$ we do find a drop in energy but the local BOO does not show any
 crystalline ordering . Interestingly the \textbf{bcc zone} for $x_A=0.50$ almost overlaps with the no crystal zone for $x_A=0.80$.  }

\label{composition_phase_plot}
\end{figure}

Formation of various crystalline forms or lack of it has been summarized in  Fig-\ref{composition_phase_plot} for various compositions and $s$ values. 
Our results agree well with the study of Vlot {\it et al.} for the equimolar mixture ($x_A=0.50$) for all $s$ values \cite{vlot}. For $0.7\leq s \leq0.74$ the systems form NaCl 
type of crystal (interpenetrating fcc)  where 
the A particles show sharp fcc peak obtained from  $q_4-q_6$ calculation. However  when we compare the local BOO for all of these systems  
we find  that as we increase the $s$ the distribution becomes broader suggesting crystalline frustration. At $s =0.76$ the system shows a sharp 
jump to bcc/sc  (all particle/ A-A pairs) crystal form. As we increase the $s$ value till $s=0.90$ this bcc/sc  (all particles/A-A pairs) signature continues. 
We refer to the region $0.76\leq$s$\leq0.90$ as the \textbf{bcc zone}.
For $s=0.92$ 
we find that the system makes a sharp transition to all atom disordered fcc + hcp form. This signature is also there for $s=0.94$. Small size disparity between A and B type of 
particles leads to a  fcc and hcp type of mixed crystal formation. Note that as the activation energy between fcc and hcp type of 
crystals is very less and there packing fraction is similar (0.74), even in single component system there is a chance of getting fcc-hcp mixed crystal \cite{frenkel}.

 Note that there is a small shift in the crystal range from that observed by Vlot. et. al \cite{vlot}. This we believe is due to the fact that unlike their system where
$\sigma_{11} =\sigma_{22}$ in our system $\sigma_{22}$ is less than $\sigma_{11}$. Our crystal structures are consistent with the lattice energy calculation of Fernandez and 
Harrowell \cite{harrowell-jcp}. Although we claim that all systems in the equimolar mixture forms crystals but there can always be some $s$ values in the transition
region for which the system will not undergo crystallization as has been observed earlier for $s=0.75$ \cite{harrowell-jcp}. 

 For the range where NaCl crystal was formed for $x_A=0.50$, we now find that for $x_A=0.80$
the A particles are arranged in a fcc lattice. The B particles although
do not show any fcc ordering as would have been expected for NaCl type crystal but the local BOO of the A particles around the B particles show sc characteristic similar to that found 
for NaCl type crystal. Also the lattice spacing between the A particles are larger than that found for pure fcc crystal and similar to that found for NaCl type structure. 
Thus we believe the absence of fcc ordering between the B particles is only due to the fact that they are lesser in number and scattered over the full system. We would assume
that this system has NaCl + fcc type crystal ordering with some defects. As reported by Fernandez and Harrowell for the equimolar mixture we too find that it is difficult to crystallize the 
system for $s>0.9$ \cite{harrowell-jcp}. However, for $x_A=0.80$ and for $s\geq0.92$ the systems easily crystallized to fcc+hcp structure. 
The local BOO parameters for $x_A=0.50$ is found to be broader 
in distribution when compared to that for $x_A=0.80$. 
This led us to conclude that for $s>0.9$ equimolar mixture ($x_A=0.50$) is more frustrated. In our study 
we also found that the drop in the enthalpy at crystallization is directly related to the width of the distribution of the local BOO and thus to frustration. 
The larger the drop the narrower is the local BOO distribution. 

However, the most interesting result here is that for  $x_A=0.80$
 we do not find any crystallization for $0.74\leq$s$\leq0.90$ (Fig-\ref{composition_phase_plot}). Although for s$=0.74$ and s$=0.90$ we do find a 
drop in energy but the local BOO does not show any crystal ordering.
Note that this range of $s$ ,except for s$=0.74$, exactly coincides with the appearance of the CsCl crystal for $x_A=0.50$.
As reported by Fernandez and Harrowell in this range the lowest energy state is a combination of CsCl+fcc crystal structure \cite{ harowell,harrowell-jcp}. 
This is quite expected because as we increase the number of A particles 
the excess A particles would like to form fcc type crystal and the AB mixture would like to form CsCl (bcc) type crystal, similar to that found for the range $0.7\leq$s$\leq0.74$ 
where
NaCl+fcc lattice is formed. However the inability of these systems to crystallize led us to believe that the stability of the supercooled 
liquid in this range of $s$ is related to the difficulty of nucleation of bcc type crystals. This difficulty in nucleation can be due to the frustration between fcc and bcc 
crystal formation as has been suggested by Fernandez and Harrowell \cite{harowell} or it can due to frustration between the LPS and both fcc and bcc crystal structures. 
In order to understand this in 
greater detail we have further studied one of the systems, $s=0.8$, in the range $0.5 \leq x_A\leq 0.9$. Note that  $x_A =0.8$ and $s=0.8$ 
represents the well known KA model.

According to the lattice energy study as we change the composition and increase the number of A type particles the lowest energy state of the system is expected to have
 a mixture of pure A fcc type crystal 
and mixed CsCl type crystal \cite{harrowell-jcp}. Since the local BOOs (both $q_{4}$ and $q_{6}$) for fcc and bcc type crystals have very similar values (see Table-1) 
it becomes difficult to identify 
the presence of both the structures unless the ordering is sharply peaked at the respective local BOO values. However, if we consider only the A particles, then they are
 expected to have both fcc and sc 
ordering if the system is a mixture of fcc and CsCl type of crystal. The local BOO of the fcc and sc are well separated, particularly in the $q_{4}$ value (see Table-1). 
Thus monitoring the A-A local BOO parameters enables us to observe the signature of both the crystalline 
forms in one system and also the transition from one form to another across the systems.    
Similar to that observed for $x_A=0.5$, the $x_A=0.55$  (for $\sigma_{ij}=0.8$) system also undergoes a CsCl type of crystallization.

\begin{figure}[h]
\centering
\begin{subfigure}{.4\textwidth}
\includegraphics[width=\textwidth]{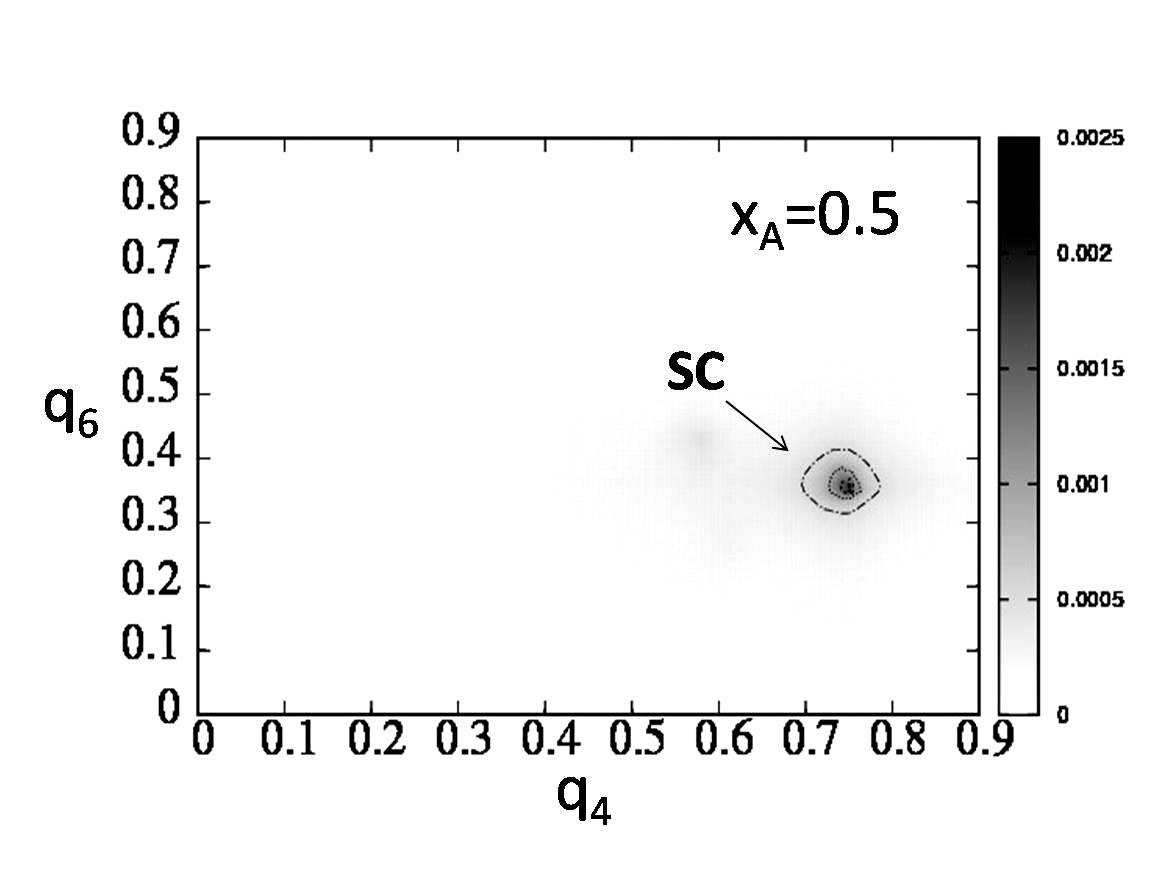}
\caption{}
\label{50_50_contour}
\end{subfigure}
\centering
\begin{subfigure}{.4\textwidth}
\includegraphics[width=\textwidth]{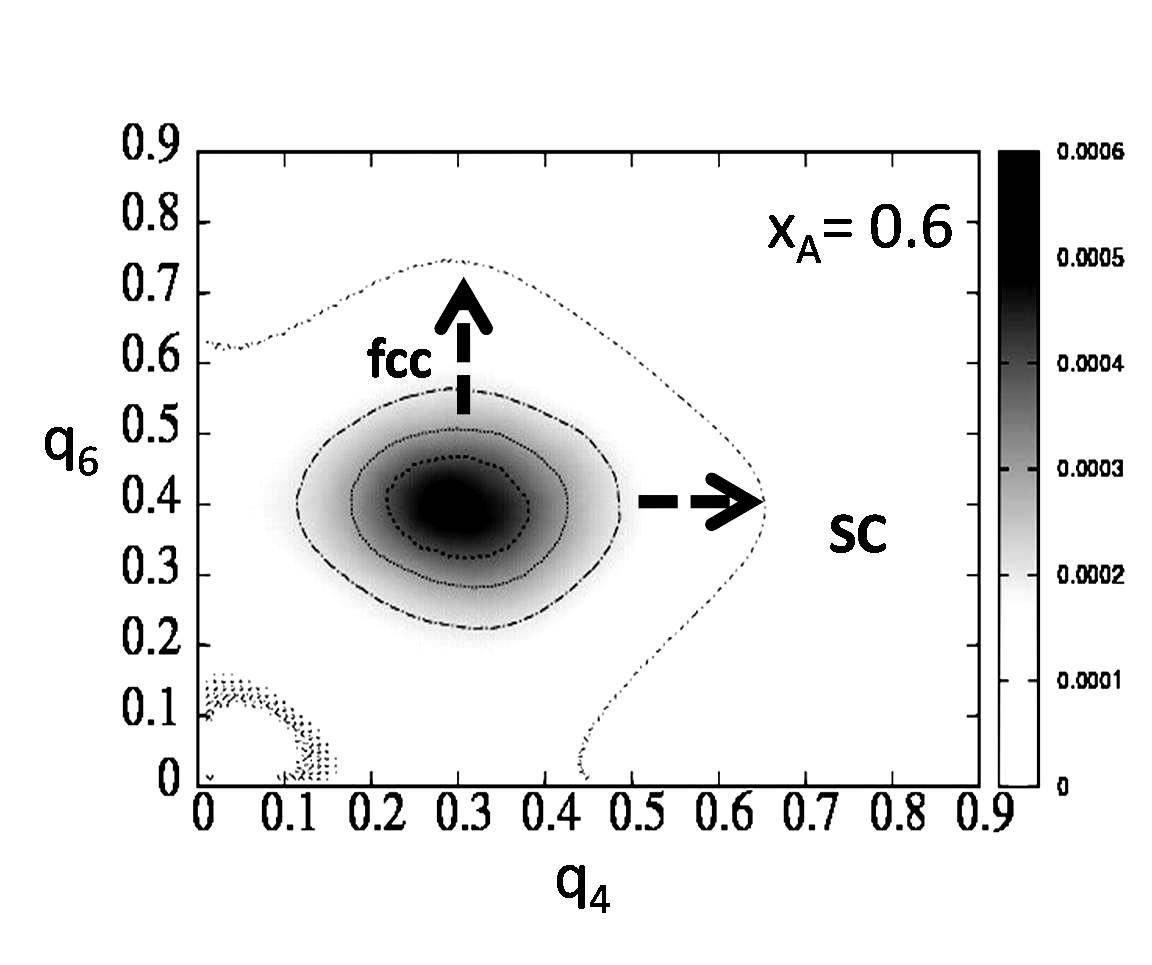}
\caption{}
\label{60_40_contour}
\end{subfigure}
\begin{subfigure}{.4\textwidth}
\centering
\includegraphics[width=\textwidth]{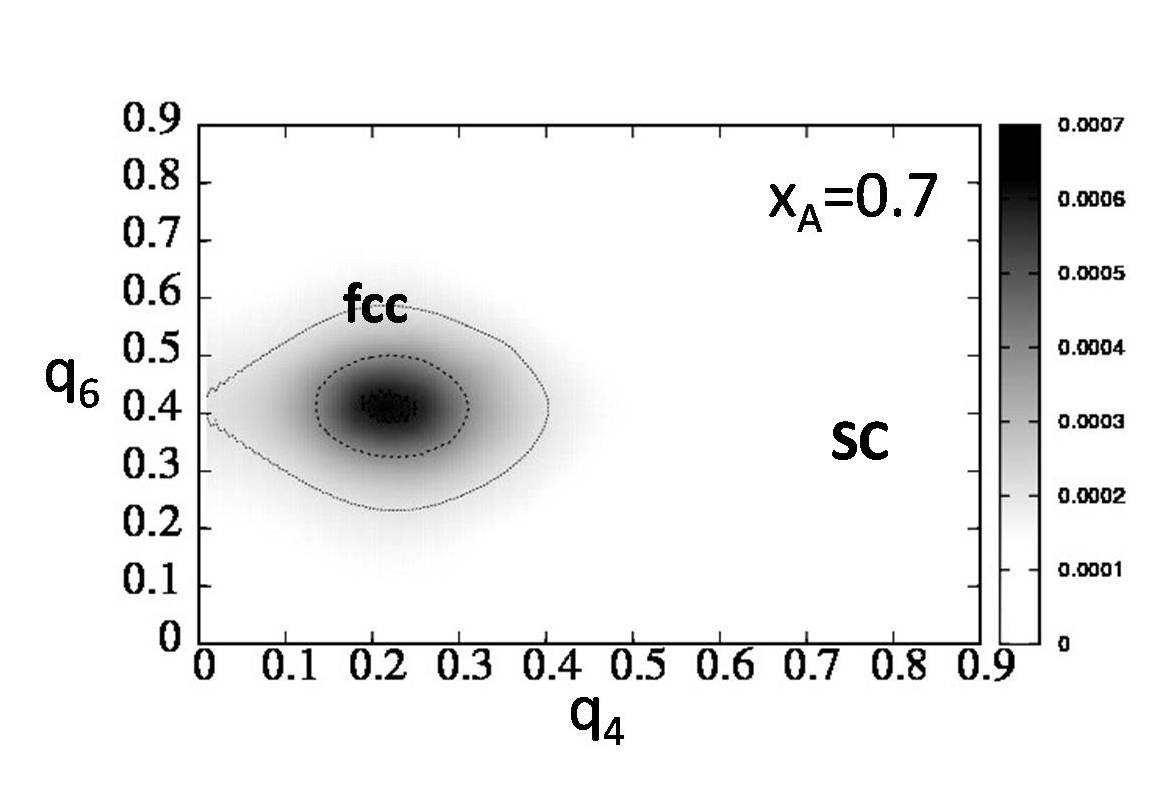}
\caption{ }
\label{70_30_contour}
\end{subfigure}
\begin{subfigure}{.4\textwidth}
\centering
\includegraphics[width=\textwidth]{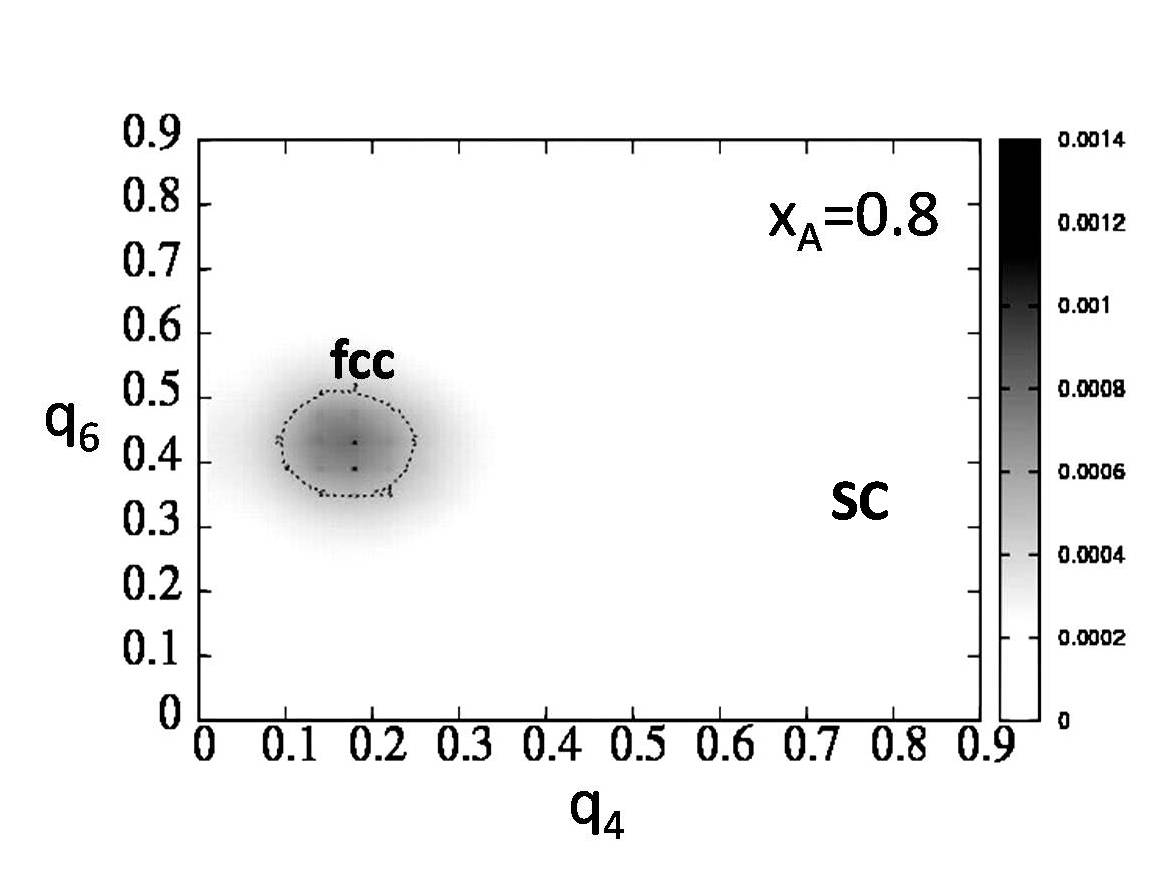}
\caption{ }
\label{80_20_contour}
\end{subfigure}
\begin{subfigure}{.4\textwidth}
\centering
\includegraphics[width=\textwidth]{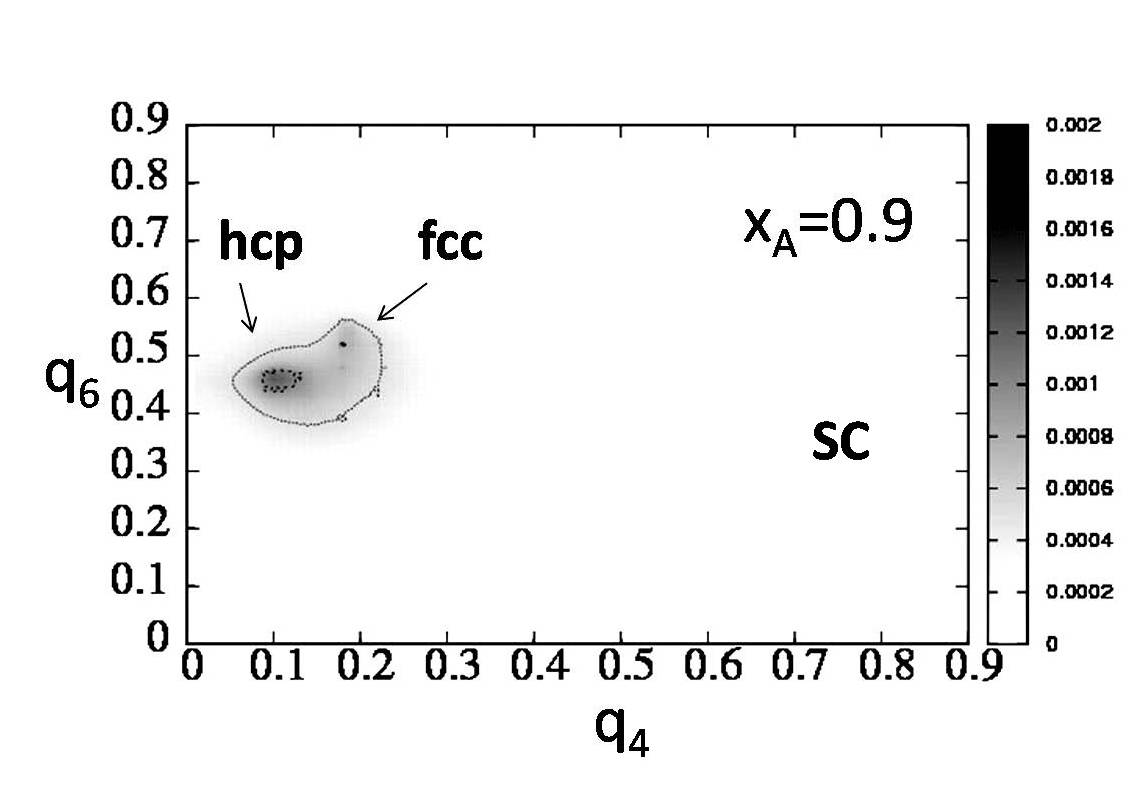}
\caption{}
\label{90_10_contour}
\end{subfigure}
\caption{ Population vs $q_4-q_6$ plot for different composition variation for $s=0.8$. If the total system is a mixture of CsCl +fcc crystalline form  then the A particles are expected
to form sc + fcc ordering.  The sc ordering of the A particles is related to the CsCl formation of the AB mixture. 
 The $q_4$ values for the fcc and sc structures are well separated (see Table-1). Thus monitoring the A-A local BOO parameters enable us to
 observe the signature of both the crystalline form in one system and also the transition from one form to the other across the systems.
 (a) For equimolar mixture ($x_A =0.5$),the distribution of population of $q_4-q_6$ is at sc position. (b) For  $x_A =0.6$ it shows a tendency towards two different forms of
 crystal structures.  Bold dotted arrows stress the ordering tendency. (c) For the composition of $x_A =0.7$, there is no tendency towards sc type of crystal formation and
there is weak tendency towards fcc type of crystal form (d) At  $x_A=0.8$ composition the system follows same trend as that for
$x_A =0.7$. (e) At $x_A =0.9$ the distribution of population of $q_4-q_6$ is at fcc and hcp position. }
\label{contour}
\end{figure}

   For the composition of $x_A =0.6$, crystallization has not been observed. We have simulated a five times bigger system to rule out any system size dependence 
and also used parallel tempering method \cite{parallel-tempering}. But the system did not undergo crystallization in any of these cases, even for a trajectory length of 10$\mu s$.
Leocmach and Tanaka have shown that the distribution of the local BOO in a liquid at low temperature can also provide information about the tendency 
of the system to undergo a certain form of crystallization \cite{tanaka-nature12}.
 As mentioned earlier to identify the pure A fcc and mixed AB bcc signature we have studied the local BOO of the A-A pairs.
The population vs $q_4-q_6$ contour plot (Fig-\ref{60_40_contour}) shows  
 a clear tendency towards both sc (bcc in total)and fcc positions for A-type of particles. Thus 
 interestingly the system with  $x_A =0.6$ although does not undergo crystallization the local BOO parameters show a strong tendency towards two different forms of crystal structures. 

For the composition of $x_A =0.7$, there is no tendency towards sc type of crystal formation for A-type of particles and  a slight tendency is there towards the fcc position 
( Fig- \ref{70_30_contour}). This is expected because the mixture now has more A-particles. For  $x_A=0.8$ the same trend follows (Fig-\ref{80_20_contour}). 
The system with the composition of $x_A =0.9$ does show crystallization of the A particles in fcc +hcp form (Fig-\ref{90_10_contour}).
 Although these results are similar to that observed by Valdes {\it et al.} \cite {valdes-jcp}, and 
Fernandez and Harrowell \cite{harrowell-jcp}, however the dual tendency for $x_A =0.6$ has not been observed earlier. 

\begin{figure}[h]
\centering
\begin{subfigure}{.76\textwidth}
\includegraphics[width=\textwidth]{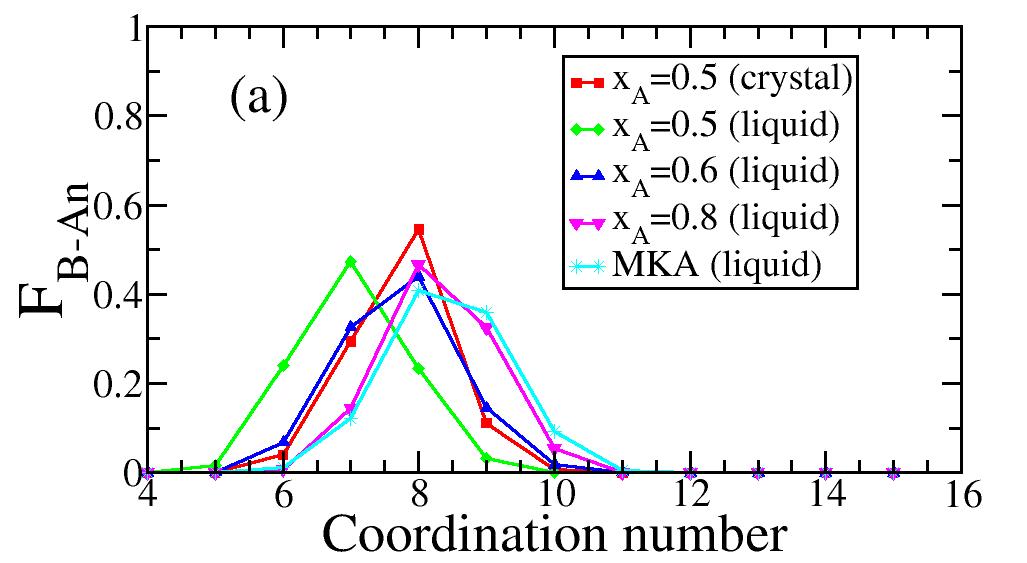}
\label{ab_0.80_CN}
\end{subfigure}
\begin{subfigure}{.8\textwidth}
\centering
\includegraphics[width=\textwidth]{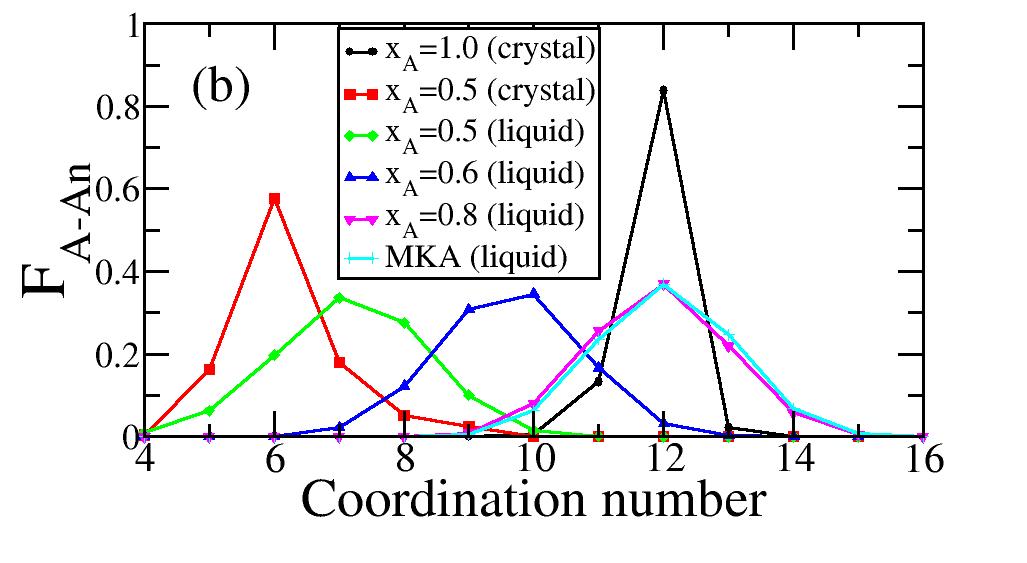}
\label{aa_0.80_CN}
\end{subfigure}
\caption{  $F_{B-An}$ describes the fraction of B particles having `n' A neighbours and $F_{A-An}$ describes the fraction of A particles having `n' A neighbours 
where n is the coordination number. (a) We have plotted $F_{B-An}$ vs n for different compositions for  $s=0.80$. 
For pure bcc crystalline form $F_{B-A8}$ should be  1. Crystal structure obtained for $x_A=0.50$ shows the peak at $F_{B-A8}$, but the liquid state of this composition shows 
the peak around $F_{B-A7}$ as obtained in Ref-9. For  $x_A=0.60$ the peak value of $F_{B-An}$ is at n=8.
(b) We have plotted $F_{A-An}$ vs n for different compositions. For pure fcc crystal 
structure $F_{A-A12}$ should be 1, and for pure bcc crystal structure $F_{A-A6}$ should be 1. In case of $x_A =0.6$ the peak value of $F_{A-An}$ is at 8, it does not satisfy any 
of these conditions. Thus the LPS does not allow the formation of either CsCl type of crystal between AB particles and fcc type of crystal between AA particles.
For the composition of $x_A=0.80$, the peak value of $F_{A-An}$ is further away from n=6 value. }
\label{coordination_number_0.80}
\end{figure}

In order to understand the origin and the effect of the dual tendency of the liquid we further analyse these systems. Both the coordination number (CN) 
 and the local BOO parameter can give us information about the locally preferred structure \cite{harrowell-jpcb}. In Fig-\ref{coordination_number_0.80} 
we plot the fraction of B particles having `n' (n=1-12) A type neighbours,
($F_{B-An}$) and 
fraction of A particles having `n' A type neighbours, ($F_{A-An}$)  at different compositions. 

For a perfect mixed bcc crystal, CsCl type, the ideal values of the parameters should be, $F_{B-A8}=1$ and $F_{A-A6}=1$. 
 We find that for  $x_A=0.50$   although there is a distribution of the parameters in the crystalline state, the peak lies at $F_{B-A8}$ and $F_{A-A6}$. The peak value of the 
parameters in the liquid state for $x_A=0.50$ does not match their crystalline values.   
 We find that $F_{B-An}$ has a peak at smaller n 
and $F_{A-An}$ has a peak at larger n when compared to its crystalline counterpart (Fig-3a, Fig-3b). As observed by Fernandez and Harrowell for a liquid to 
form CsCl type crystal the $F_{B-An}$ 
is always found to be lower than the ideal value \cite{harrowell-jpcb}. This lower value might help the rearrangement of the neighbours between the neighbouring A and B particles to form a bcc seed 
where the A particle can give away 
one of its extra neighbour to the neighbouring B particle. In case of  $x_A=0.80$ we find that 
the $F_{B-An}$ peak has shifted to n=8 but at the same time the $F_{A-An}$ peak has shifted to n=8 which is away from the 
n=6 value required for a CsCl structure or n=12 required for the fcc structure. Thus even if the B particles have required neighbours the surrounding A particles have more
 A neighbours 
than that required for the formation of the CsCl type crystal. It will require large rearrangement of neighbours between the A particles to form bimodal distribution of 
its neighbours with peaks at n=6  and n=12. Thus this locally preferred structure does not allow the formation of either CsCl type crystal between AB particles or 
fcc crystal between the AA particles. For the  composition of $x_A=0.80$ the $F_{A-An}$ distribution moves further away from the n=6 value. Hence we can say that there is a frustration 
between LPS, and the global structure which is a combination of  bcc and fcc crystal structure.

\begin{figure}[h]
\centering
\begin{subfigure}{0.72\textwidth}
\includegraphics[width=\textwidth]{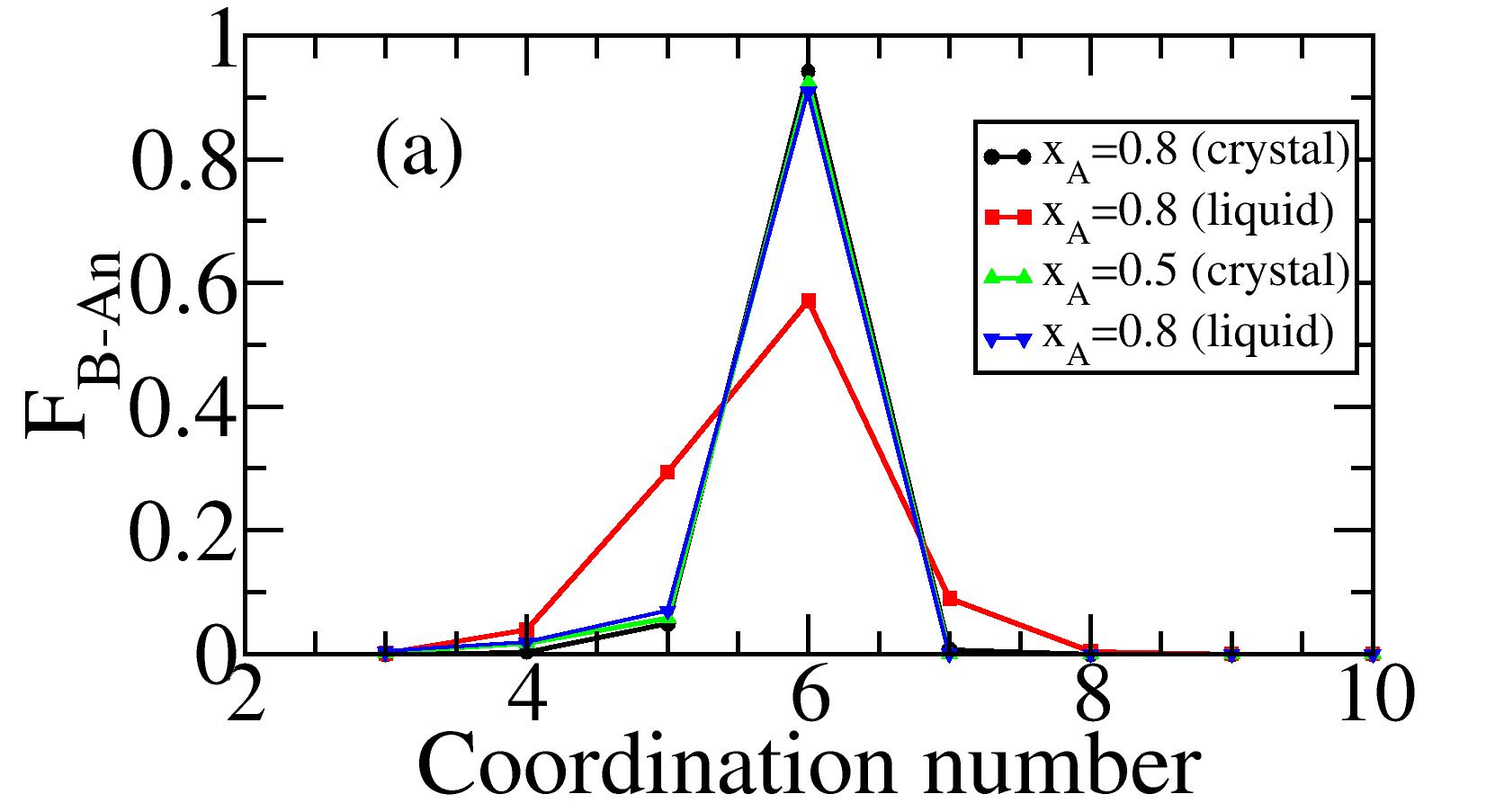}
\label{ab_0.70_CN}
\end{subfigure}
\begin{subfigure}{0.8\textwidth}
\centering
\includegraphics[width=\textwidth]{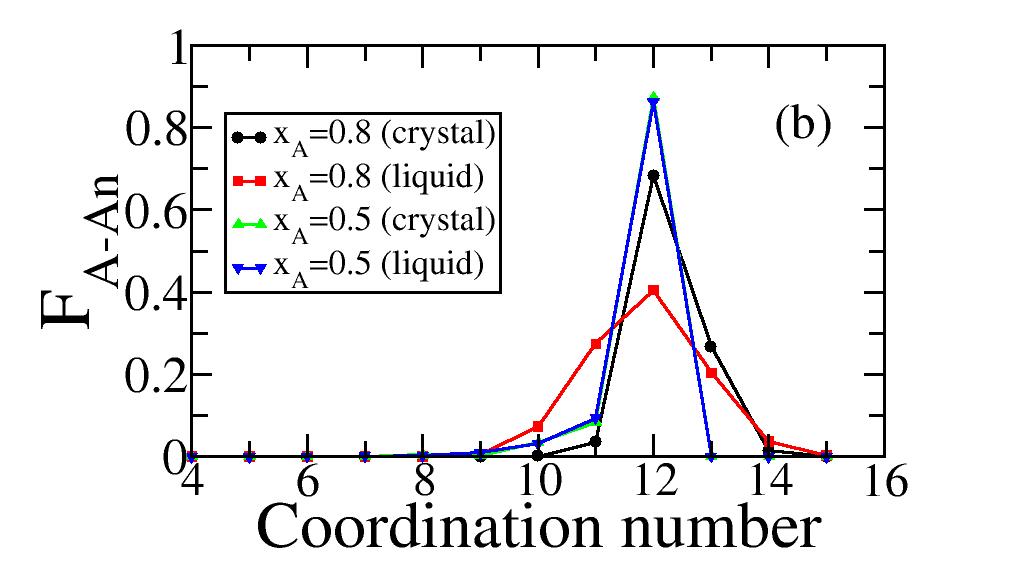}
\label{aa_0.70_CN}
\end{subfigure}
\caption{ $F_{B-An}$ and $F_{A-An}$ are the same as defined in Fig-\ref{coordination_number_0.80}. 
 (a) We have plotted $F_{B-An}$ vs n for different compositions for  $s=0.70$ .  
For both NaCl and  NaCl +fcc type crystal structure $F_{B-A6}$ should be 1. Here the plots for the crystal structure of $x_A=0.8$, liquid and crystal structures of $x_A=0.5$
are overlapping. 
 (b) We have plotted $F_{A-An}$ vs n for different compositions for the same s value.
For both NaCl and NaCl + fcc type crystal $F_{B-A6}$ should be 1. Here we find that the peak positions are at their  expected crystalline values. 
So there is less frustration between LPS and NaCl + fcc form for $x_A=0.80$. }
\label{coordination_number_0.70}
\end{figure}

In order to further substantiate our claim and also to show that the change in coordination number between A-A particles is not only a density (of A particles)
 effect, we do the same coordination number analysis for the system with $s=0.7$ value for two compositions. 
This system is known to crystallize 
in NaCl form for the composition of $x_A=0.50$ and NaCl+fcc 
form for the composition of $x_A=0.80$. We plot the $F_{B-An}$ values and $F_{A-An}$ values for both the systems in their liquid and crystalline states (Fig-4a, Fig-4b).
 We notice that in all the cases the 
$F_{B-An}$ peaks at n=6 and $F_{A-An}$ peaks at n=12. This should be true for all the other intermediate compositions also. Since for both NaCl and fcc crystal the $F_{A-An}$ 
needs to peak at n=12 
thus the coordination number around a A particle does not require much rearrangement for the system to crystallize. This shows that there is less frustration 
between the LPS and the  NaCl+fcc crystal.  
This is precisely where the CsCl+fcc and NaCl+fcc crystal differ from each other. 
Thus we can infer that due to the requirement of large rearrangement of neighbours between the A-A particles the bcc crystallization has a large nucleation barrier. 
Our analysis further reveals that this should be true not only for the KA model but for any system which is in the bcc zone and has large value of $x_A$. 
The coordination number analysis shows that this barrier for crystallization to the bcc type structure should become 
higher as we increase the composition of the A particles. 
Our picture of frustration is similar to that given by Tanaka and co-workers  who
 claim that when there is a mismatch between the LPS and the global structure, the LPS acts as a source of frustration against crystallization 
\cite{ tanaka-nature06,tanaka-vshaped-prl, tanaka22, tanaka23, tanaka29}.

However, although the barrier for bcc crystallization increases as the composition is increased but the tendency for fcc formation also increases at the same time. 
For the composition of $x_A=0.80$ both the local BOO and 
the $F_{A-An}$ distribution show fcc like characteristic. Toxvaerd {\it et al.} have modified the KA model (MKA) by reducing the A-B attraction parameter \cite{dyre}. 
They have claimed that this modified model 
can help to predict the crystallization process of the KA model. The parameters for the MKA model in its liquid state has been plotted in Fig 3. It indeed has a resemblance 
with that of the KA model and the MKA model is reported to show a crystallization of the A particles .

We also observe that around the \textbf{bcc zone} there are two different types of fcc crystal, one is the disordered fcc crystal at higher $s$ value  and the other is  the NaCl+fcc crystal at lower $s$ value .
 The KA model shows a tendency towards crystallizing in fcc type structure. This tendency should be present for the whole \textbf{bcc zone} in the range of $0.70\leq x_A < 0.9$.

\begin{figure}[h]
\centering
\includegraphics[width=.8\textwidth]{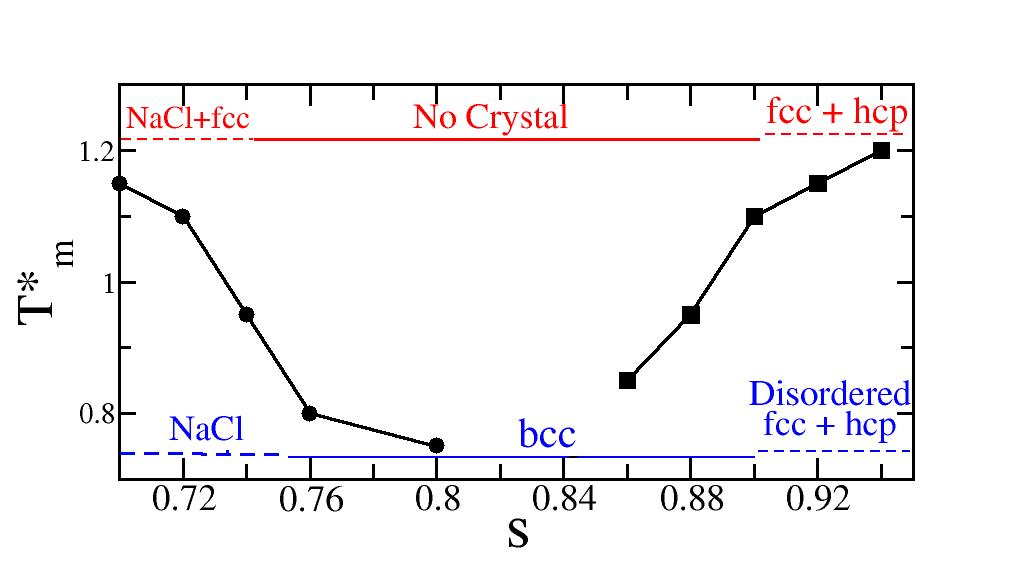}
\caption{ V-shaped phase diagram of two different variants of fcc crystal structure.  Melting points for Nacl + fcc type of crystal (black solid circles) and
 mixed hcp +fcc crystal form (black solid squares) for different $s$ values are plotted here  \cite{melting}. 
We donot find any triple point. Here blue lines (dotted and solid)  denote the range where various types of crystal forms are found for  $x_A=0.50$. Red dotted lines  
denote the same for $x_A=0.80$. We do not see any crystallization in the range shown by the red solid line.  Black solid lines are guide to the eyes.} 

\label{phase_plot_temp}
\end{figure}

Thus it is imperative to understand the tendency of crystallization of the \textbf{bcc zone} to form any of these two types of fcc crystal.
 In order to do that we study the melting of the disordered fcc crystal (formed for $s=0.94$) 
 and NaCl+fcc 
crystal (formed for $s=0.7$)  by varying the $s$ value.  We start with the crystalline structures obtained for the system with $s=0.7$ and $s=0.94$ for $x_A=0.80$.
 The systems are first cooled to $T^*=0.1$ and then these structures are taken as a reference structure for different $s$ values \cite{melting}. It is interesting to note that the local BOO parameter  
obtained for the NaCl+fcc structure with $s=0.8$ is similar to that obtained for the MKA model of Toxvaerd {\it et al.} \cite{dyre}. This shows that the NaCl+fcc structure is where the MKA model 
crystallizes and the KA is expected to crystallize.   The $T-s$ phase diagram clearly shows that the stability of any form of fcc type crystal is less in the \textbf{bcc zone}. 
In our study we could not predict a triple point as in the range  $0.80<s<0.86$ none of the crystal forms were found to be stable even at $T^*=0.1$ (Fig-\ref{phase_plot_temp}). 
The energy per particle at T=0 and P=0 for
 the KA model in this NaCl+fcc structure was found to be -7.291 which is higher than the energy per particle for the amorphous state reported earlier (-7.72) 
\cite{harowell,srikanth}. Thus it might be possible that the KA model will never crystallize even in the NaCl+fcc form. 
The phase diagram found here is similar to that obtained by Molinero {\it et al.} for Si-like potential by modifying the tetrahedral character in 
the Stillinger-Weber potential \cite{valeria} and by Tanaka { \it et al.} for water-LiCl mixture \cite{tanaka-vshaped-prl}. According to Tanaka this kind of V shaped diagram is
 related to the Glass forming ability of the system where systems sitting at the bottom of the V have higher GFA and are stable against crystallization
  \cite{tanaka-epje,tanaka-vshaped-prl}.

 \section{Conclusion}

In this article we have tried to understand the interplay between the crystallization and the glass transition in binary Lennard-Jones mixtures.
 The study has explored the effect of the inter-species interaction length ($s$) and also the composition.  The systems studied here have  
negative enthalpy of mixing and the size ratio between the components are always kept 12\%. For the range of $s$ studied here the equimolar mixture crystallizes 
into three 
different forms of crystals similar to that found by Vlot { \it et al.} \cite{vlot}. For the large and the small $s$ values distorted fcc + hcp structure
 and interpenetrating 
fcc structure (NaCl type) are found respectively.
 The systems with intermediate $s$ values are found to form bcc structure (CsCl type). For $x_A=0.80$ although the systems with small and 
large $s$  values crystallize to NaCl+fcc and hcp + fcc crystal, respectively, the bcc zone does not crystallize. This shows that the frustration against
 crystallization has a connection with the formation of bcc crystal structure.  
The study with $s=0.80$ at different compositions gives further insight to this frustration. The LPS of the composition for $x_A=0.60$  analysed
 from local BOO and coordination number show a strong frustration between LPS and both bcc and fcc crystal forms. The LPS does not favour either of the crystal form. However the LPS 
 for $x_A=0.80$ shows a tendency towards fcc crystal formation.
 Thus we can claim that as the composition of A particle increases, the nucleation barrier to form a bcc crystal also increases. This conclusion is coherent with the finding of 
Fernandez and Harrowell.
They have reported that even after putting a bcc seed in a the KA mixture they have not found the growth of bcc crystal \cite{harrowell-jcp}.
 This must also be the reason why Toxvaerd {\it et al.} could form a mixed fcc + bcc phase in the KA mixture only
after putting the complete bcc structure and allowing the growth of fcc lattice around it \cite{dyre}. However, Valdes {\it et al.} have reported that for $x_A=0.30$ the low temp state of
the system seems to be composed of bcc+fcc crystalline structure \cite{valdes-jcp}. This shows that instead of increasing the bigger particles if we increase the composition of
smaller particles  then nucleation barrier for bcc crystalline structure will reduce. 
It will be interesting to perform a free energy calculation of the nucleation barrier for bcc crystal formation at different compositions similar to that performed for other
systems \cite{monson, swetlana}.
This is beyond the scope of the present work and would be taken as a future project. 

 Since the LPS for  $x_A=0.80$  shows a fcc characteristic, we have also studied the phase diagram of the melting temperature of two different fcc types 
of crystal forms which are present for the higher and lower $s$ values. The study shows that
the phase diagram has a V-shape where the bcc zone which does not crystallize sits at the bottom of the V. This V-shaped phase diagram has also been 
observed earlier for Si-like system and also for water-LiCl system. It has been always found that due 
to frustration between the LPS and the global structure, the systems sitting at the bottom of the V are good glass formers. 

 Although we have not studied the phase diagram by varying the composition , but the local BOO and CN analysis predicts a similar V-shaped phase diagram where at $x_A=0.50$
  the system forms bcc type crystal  and the pure monoatomic
system ($x_A=1.0$) forms fcc type crystal. For the intermediate values of $x_A$ where the crystal structure analysis shows that the mixture of fcc+CsCl is the global structure,
 the analysis of the LPS shows that there is a frustration between
the LPS and the global structure. Thus the picture suggests that the intermediate $x_A$ values will be sitting at the bottom of the V and the $x_A=0.5$ and $x_A=1.0$ will
 be forming the two ends. Hence the bcc zone for composition of $x_A=0.80$ is a good
glass former not only due to the frustration between the two different fcc lattice structures  but also due to the frustration between the LPS and fcc+bcc lattice formation. 

Our study suggests that whenever we increase the composition of one of the species of a binary system which in its equimolar composition forms bcc crystal (CsCl type) we 
will find a frustration between the LPS and global structure.
In more general terms if a global structure of a mixed system has two crystalline forms such that any of the species which is present in both the
 crystal structures has a large difference in its order parameter (coordination number or local BOO or any 
other order parameter) in the two crystal forms, there will be frustration between the LPS and the global structure.
The LPS will not be closer to either of the crystalline states
 and this frustration will lead to the stability of the system against crystallization.
   
  \section{Acknowledgements}

This work has been supported by the Department of Science and Technology (DST), India and CSIR-Multi-Scale Simulation and Modeling project.
 AB thanks DST for fellowship. Authors thank Dr. Srikanth Sastry, 
Dr. Rahul Banerjee, Dr. G. Kumaraswamy, Dr. Mantu Santra, Dr. Vishwas Vasisht, Mr. Rajib Biswas
 for discussions.

\clearpage


\begin{thebibliography}{10}
\bibitem{melting}
The two reference structures used for the melting temperature study are
  obtained by cooling the systems ($s=0.70$ and $s=0.94$ for $x_{A}=0.8$ ) to
  ${T}^*=0.1$ . {T}he structures thus obtained at $s=0.70$ is a {N}a{C}l +fcc
  mixture and at $s=0.94$ is fcc+hcp. {T}hese reference structures at different
  $s$ values are stepwise heated (0.05 {T}* every 10 ns) to obtain the melting
  temperature, ${T}^*_m$.
\end{thebibliography}

 
\clearpage
 
\end{document}